\documentclass[aps,prl,twocolumn,showpacs,superscriptaddress,twoside,showpacs]{revtex4-2}
\usepackage{graphicx}
\usepackage{dcolumn}
\usepackage{bm}
\usepackage{booktabs}
\usepackage{amsmath}
\usepackage{color}

\begin{document}
	\title{Comment on ``Longitudinal wobbling in $^{133}$La [Eur. Phys. J. A 55, 159 (2019)]"}

\author{W. Hua}
\affiliation{Sino-French Institute of Nuclear Engineering and Technology, Sun Yat-Sen University, Zhuhai 519082, China} 
\author{S. Guo}
\thanks{{\it Corresponding author}}
\email{gs@impcas.ac.cn}
\affiliation{CAS Key Laboratory of High Precision Nuclear Spectroscopy, Institute of Modern Physics, Chinese Academy of Sciences, Lanzhou 730000, China} 
\affiliation{School of Nuclear Science and Technology, University of Chinese Academy of Science, Beijing 100049, People's Republic of China}
\author{C. M. Petrache}
\affiliation{Universit\'{e} Paris-Saclay, CNRS/IN2P3, IJCLab, 91405  Orsay, France}

\date{}

\begin{abstract}
	
	In [S. Biswas et al., Eur. Phys. J. A 55, 159 (2019)] a longitudinal wobbling band was reported in $^{133}$La. The critical experimental proof for this assignment is the E2 dominated linking transitions between the wobbling and normal bands, which are supported by angular distribution and linear polarization measurements. However, severe problems are found in the reported experimental information, indicating that the assignment of wobbling band was not firmly established.

\end{abstract}



\maketitle

The band built on the 17/2$^-$ state at 1738 keV in $^{133}$La which has been interpreted as the signature partner of the band built on the 11/2$^-$ state at 536 keV in Ref. \cite{Petrache}, has been reinterpreted as longitudinal wobbling in Ref. \cite{133La} based on angular distribution and polarization measurements.  As the first experimental candidate for low-spin longitudinal wobbling, this work is rather important, and deserves a careful examination on the reliability of the reported results.

After evaluating the reported experimental information carefully, some problems are found. To establish a wobbling band, the key experimental criterion is the $E2$ dominated $\Delta I=1$ linking transitions. In the commented work \cite{133La}, $E2/M1$ mixing ratios ($\delta$) were deduced by the angular distribution and linear polarization measurements. For three linking transitions with energies of 758, 874, and 982 keV, the mixing ratios are all deduced with absolute values larger than 1, leading to a predominantly electric nature.

\begin{figure}[ht]
	\vskip -. cm
	\hskip -. cm
	\centering\includegraphics[clip=true,width=0.48\textwidth]{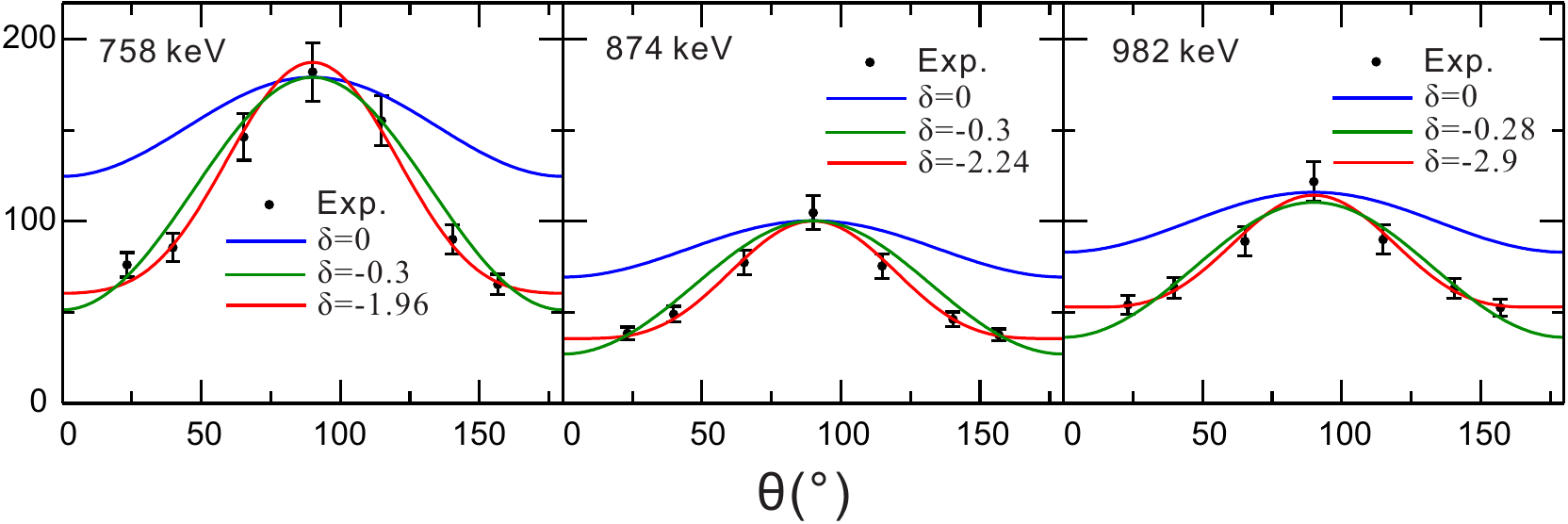}
	\vskip -0. cm
	\caption{(Color online) Estimated angular distribution curves and experimental results for the reported longitudinal wobbling band in $^{133}$La. The experimental results and the curves for pure $M1$ transition (in blue) and large $\delta$ (in red) are taken from the Fig. 2 in Ref. \cite{133La}, and those with smaller mixing ratios and same $\sigma$/I are also plotted (in green) for comparison. }
	\label{fig1}
\end{figure}

For the angular distribution, estimated curves with fitted $a_2$ and $a_4$ coefficients were shown in comparison with the experimental results, and the mixing ratios were deduced based on the  $a_2$ and $a_4$ coefficients (see Fig. 2 in Ref. \cite{133La}).
The same curves are produced in the present comment, using the data of Ref. \cite{133La} and a direct method \cite{Matta}, and adding an extra curve for each transition produced by using small mixing ratios ($|\delta|<1$) (see Fig. \ref{fig1}). It is found that the curves with larger and smaller mixing ratios are quite similar, and are hardly distinguishable within the reported error bars. It appears therefore that the dominating $M1$ character of the transitions cannot be firmly excluded based on only the angular distribution measurement.

\begin{figure}[t]
	\vskip -. cm
	\hskip -. cm
	\centering\includegraphics[clip=true,width=0.4\textwidth]{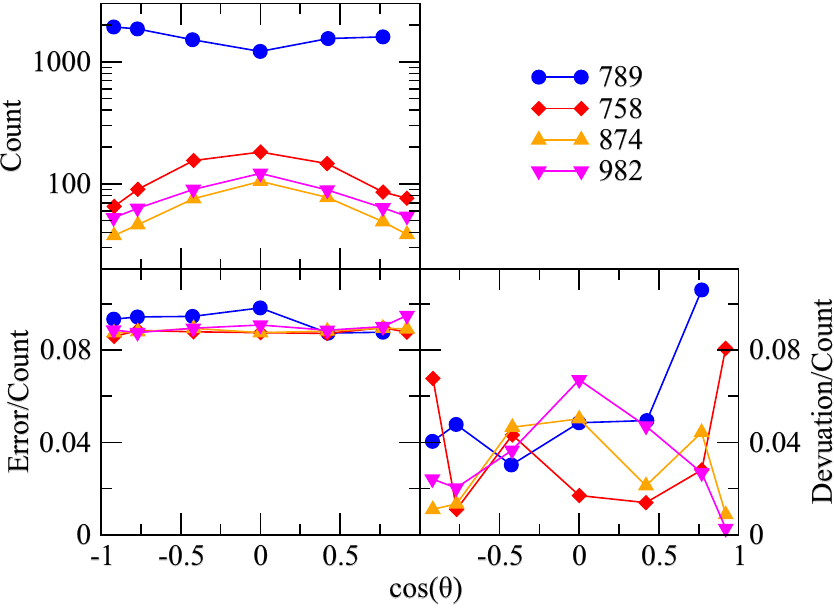}
	\vskip -0. cm
	\caption{(Color online) The counts, relative errors and relative deviations for the experimental results on the 789-, 758-, 874-, and 982-keV transitions in Ref. \cite{133La}.}
	\label{fig2}
\end{figure}

Furthermore, the relative errors of the deduced experimental results are expected to be smaller if the statistics is higher. In the commented paper, the number of counts of the 789-keV $\Delta I = 2$ pure $E2$ transition of the yrast band built on 11/2$^-$ is higher by about an order of magnitude relative to the linking transitions of the yrare band (see Fig. \ref{fig2} and Fig. 3 in Ref. \cite{133La}). However, the relative errors of the counts on the 789-keV transition are even slightly higher than those on the three linking transitions. In addition, for each transition, the number of counts at different angles is different. This is in contrast with the relative errors of the three linking transitions which are almost identical. We also checked the deviation between the experimental results and the estimated curve, which for the 789-keV transition is expected to be smaller than for the linking transitions, because its statistics is higher. Based on the errors and the deviations between the experimental and calculated angular distribution, the reported results of Ref. \cite{133La} appear to be questionable, and for the three linking transitions in particular, significantly underestimated.

\begin{figure}[b]
	\vskip -. cm
	\hskip -. cm
	\centering\includegraphics[clip=true,width=0.5\textwidth]{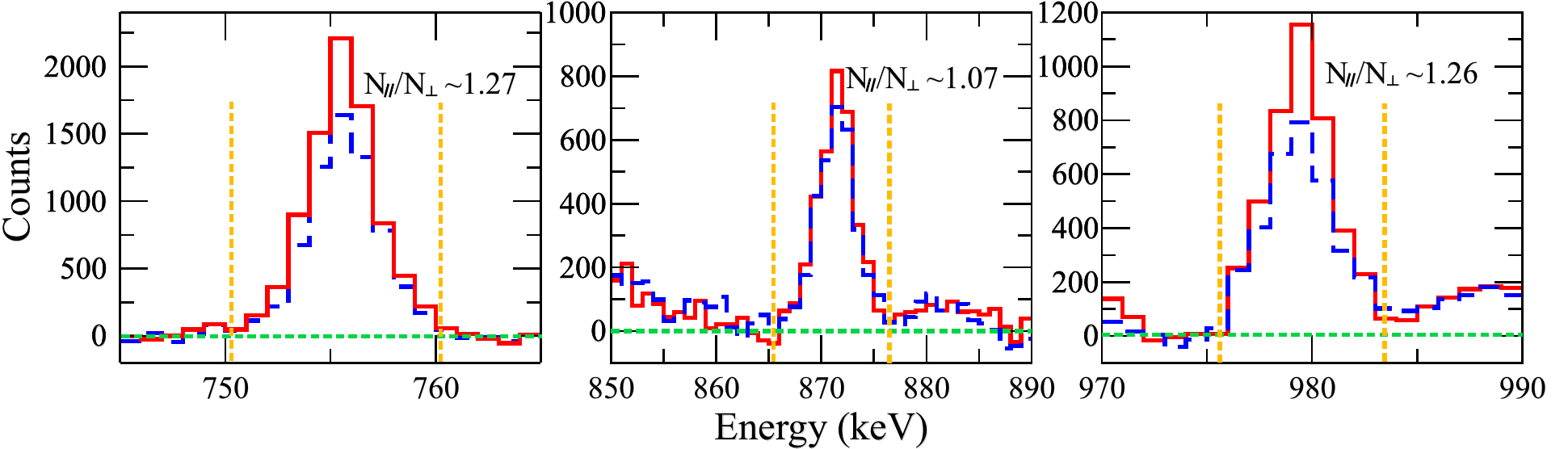}
	\vskip -0. cm
	\caption{(Color online) The spectra taken from Fig. 4 in Ref. \cite{133La}. The $R$ values are deduced within the range between the orange dash lines assuming the a zero background (green dash lines).}
	\label{fig3}
\end{figure}


Another problem exists in the extraction of the polarization values from the coincidence events between the segments of the Clover detector in the direction perpendicular ($N_\perp$) and parallel ($N_\parallel$) to the emission plane. According to Ref. \cite{PDCO},

\begin{equation}
	Q = \frac{a R-1}{a R+1}/P.\label{e1}
\end{equation}

\noindent where $Q$ denotes the polarization sensitivity of the CLOVER detectors, $P$ denotes the polarization values, $R$ denotes the ratio between $N_\perp$ and $N_\parallel$, and $a$ denotes correction due to the asymmetry in response of the Clover segments.

\begin{figure}[t]
	\vskip -. cm
	\hskip -. cm
	\centering\includegraphics[clip=true,width=0.3\textwidth]{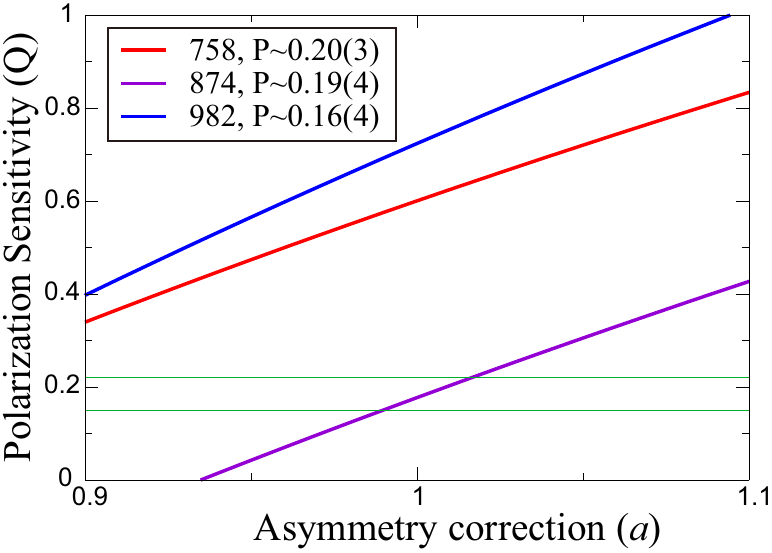}
	\vskip -0. cm
	\caption{(Color online) Each curve shows possible solutions of asymmetry correction ($a$) and polarization sensitivity ($Q$) to deduce the reported polarization values ($P$) from the reported spectra for the three linking transitions of 758, 874 and 982 keV. As a reference, the area between two green lines show the reasonable values for $Q$ according to a calibration in Ref. \cite{PDCO}.}
	\label{fig4}
\end{figure}

According to the Fig. 4 in Ref. \cite{133La}, the $R$ values are deduced assuming a zero background (see Fig. \ref{fig3}).
We also extracted the polarization values from the data in Fig. 4 of Ref. \cite{133La} and list them in Fig. \ref{fig4}.

It is found that the $R$ value for the 874-keV transition is much lower than those for the 758- and 982-keV transitions, while the $P$ values are only slightly different. According to the Eq. \ref{e1}, only two coefficients, $Q$ and $a$, are involved in the deduction of $P$. The parameter $a$ is calibrated to increase or decrease linearly, while $Q$ decreases slightly with increasing $\gamma$ energies in the 700-1000 keV interval. The curves in Fig. \ref{fig4} are plotted according to Eq. \ref{e1}. For each transition, the suitable $Q$ and $a$ values should be close to the corresponding curves. The calibration of the polarization sensitivity was not mentioned in Ref. \cite{133La}. One observes that the curve for the 874-keV transition is well below those for the 758- and 982-keV transitions. It is therefore impossible to find three points on the three curves which would induce a monotonic change of {\it a} and  $Q$ values with decreasing transition energy. Therefore it is questionable how the polarization values were deduced.

In conclusion, severe problems have been found in the reported experimental results in Ref. \cite{133La}, which sheds doubts on the existence of the reported longitudinal wobbling. Meanwhile, considering the importance of the experimental observation on the low-spin longitudinal wobbling, another experimental work and reliable data analysis would be very welcome to clarify the identified incongruences in Ref. \cite{133La}.

This work has been partly supported by the National Natural Science Foundation of China, under contracts No. 11805289 and U1932137.

\end{document}